\documentclass{ws-ijmpcs}
\let\captionbox=\undefined
\usepackage{caption}

\usepackage{subfig}
\usepackage[percent]{overpic}
\usepackage{xcolor}

\begin{document}

\markboth{Markus Oberle}
{Quasi-free photoproduction of pion-pairs off protons and neutrons}

\catchline{}{}{}{}{}
%
%%%%%%%%%%%%%%%%%%%%%%%%%%%%%%%%%%%%%%%%%%%%%%%%%%%%%%%%%%%%%%%%%%%%

\title{Quasi-free photoproduction of pion-pairs off protons and neutrons}

\author{M. Oberle and B. Krusche\\
for the A2-, Crystal Ball- and TAPS-collaborations}

\address{Department of Physics, University of Basel, Klingelbergstrasse 82\\
Ch-4056 Basel, Switzerland,
\\
Markus.Oberle@unibas.ch}

\maketitle

\begin{history}
\received{Day Month Year}
\revised{Day Month Year}
\end{history}

\begin{abstract}
Beam-helicity asymmetries and mass-differential cross sections have been measured at
the MAMI accelerator in Mainz for the photoproduction of neutral and mixed-charge pion 
pairs in the reactions $\boldsymbol{\gamma}p\rightarrow n\pi^0\pi^+$ and 
$\boldsymbol{\gamma}p\rightarrow p\pi^0\pi^0$ off free protons and $\boldsymbol{\gamma}d
\rightarrow (p)p\pi^0\pi^-$, $\boldsymbol{\gamma}d\rightarrow (n)p\pi^0\pi^0$ and 
$\boldsymbol{\gamma}d\rightarrow (n)n\pi^0\pi^+$, $\boldsymbol{\gamma}d\rightarrow (p)n
\pi^0\pi^0$ off quasi-free nucleons bound in the deuteron for incident photon energies 
up to 1.4 GeV. Circularly polarized photons were produced in bremsstrahlung processes 
of longitudinally polarized electrons and tagged with the Glasgow-Mainz magnetic spectrometer. 
The decay products (photons, protons, neutrons, and charged pions) were detected in the 
$4\pi$ electromagnetic calorimeter composed of the Crystal Ball and TAPS detectors. 
Using a full kinematic reconstruction of the final state, excellent agreement was found 
between the results for free and quasi-free protons, indicating that the quasi-free neutron 
results are also a close approximation of the free-neutron results. Comparisons of the results 
to predictions from model calculations suggest that especially the reaction mechanisms in the 
production of the mixed-charge final states are still not well understood, in particular at 
low incident photon energies in the second nucleon resonance region.

\end{abstract}

\ccode{PACS numbers: 11.25.Hf, 123.1K}

\section{Introduction}
The fundamental properties of the strong interactions in the non-perturbative range
are closely related to the excitation spectrum of nucleons, which is not fully 
understood yet and therefore represents an intensively discussed topic. The apparently poor match 
\cite{Krusche_03} between the experimental database for excited nucleon states and 
model predictions based on Quantum Chromodynamics `inspired' quark models 
has motivated huge efforts in experimental and theoretical development.
To overcome the limitations in the available database, which was dominated by the results 
from pion scattering on nucleons and thus biased against nucleon resonances with 
small couplings to $N\pi$, was the primery target. Huge advances in accelerator and 
detector technology allow for photoproduction of mesons, which thus has become a prime tool 
in this research. 
Sequential decays involving intermediate excited states are playing an essential role in these
processes and are especially important for resonances from the third resonance 
region and above, where the cross sections from many single-meson channels are decreasing. These 
types of reactions have attracted much interest in recent years and thus the photoproduction of 
pseudoscalar meson pairs, mostly $\pi\pi$ but also $\pi\eta$, has been
studied in detail experimentally (see 
%\cite{Sarantsev_08,Thoma_08,Zehr_12,Kashevarov_12,Ajaka_08,Horn_08a,Horn_08b,Kashevarov_09,Kashevarov_10} 
Refs. \cite{Sarantsev_08}$^-$\cite{Kashevarov_10}
for recent results).
The measurement of pion pairs with different charge combinations and also 
measurements of their production off both nucleons is mandatory for an isospin
decomposition of the reaction, helping to identify contributions from $N$ and 
$\Delta$ resonances. 
At least 23 independent variables would have to be measured in
order to fix amplitudes 
and phases in the description of photoproduction of pseudo-scalar meson pairs \cite{Roberts_05} 
and thus a complete measurement appears unrealistic. But already the measurement of at 
least some polarization observables can provide valuable constraints for the reaction models, 
since these observables are often very sensitive to small details in the reaction mechanisms.

\section{Experiment and analysis}

The experiments were performed at the tagged photon facility at the Mainz
Microtron accelerator 
MAMI\cite{Herminghaus_83,Kaiser_08}. Detailed information about the different components of
the detector setup and analysis techniques can be found in 
%\cite{Anthony_91,Hall_96,McGeorge_08,Starostin_01,Novotny_91,Gabler_94,Schumann_10,Watts_04,Zehr_12}.
Refs. \cite{Zehr_12}$^,$\cite{Anthony_91}$^-$\cite{Watts_04}.
%\vspace{-3mm}
\begin{figure}[!ht]
\centering{  \subfloat{%
    \begin{overpic}[width=0.65\textwidth]{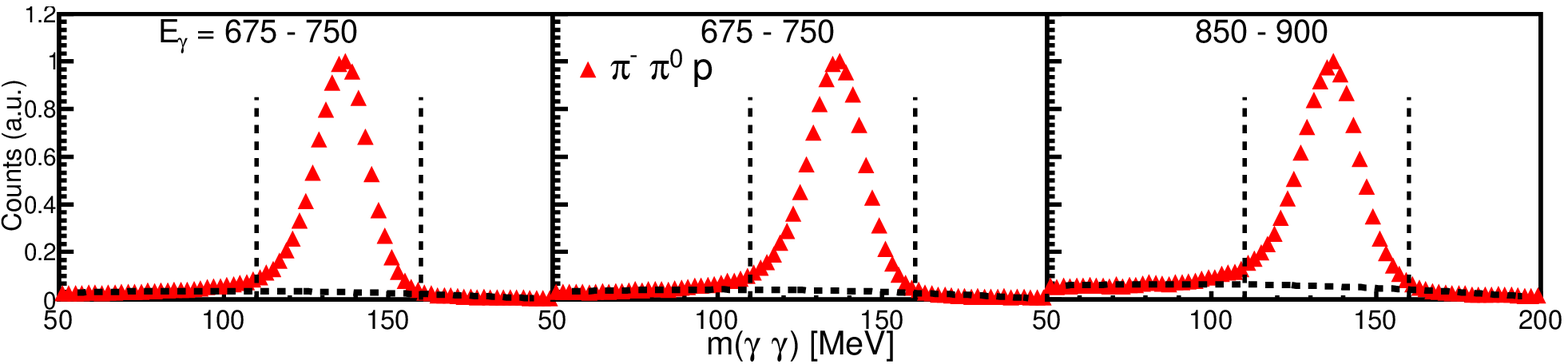}
      \put(40,3){{\color{green} \large{\rotatebox{45}{Preliminary}}}}
    \end{overpic}
  }}\newline
\centering{    \subfloat{%
    \begin{overpic}[width=0.65\textwidth]{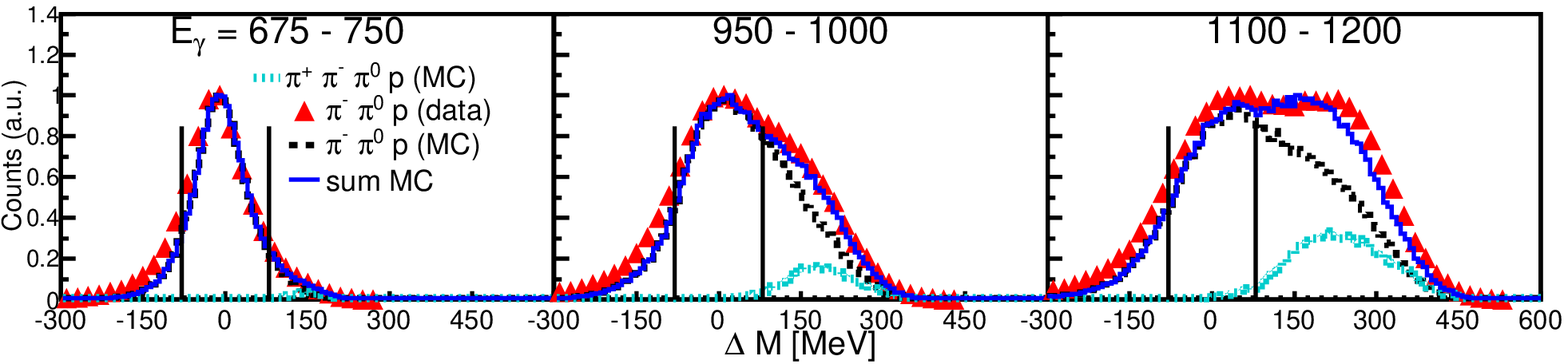}
      \put(40,3){{\color{green} \large{\rotatebox{45}{Preliminary}}}}
    \end{overpic}
  }}\newline
\centering{    \subfloat{%
    \begin{overpic}[width=0.65\textwidth]{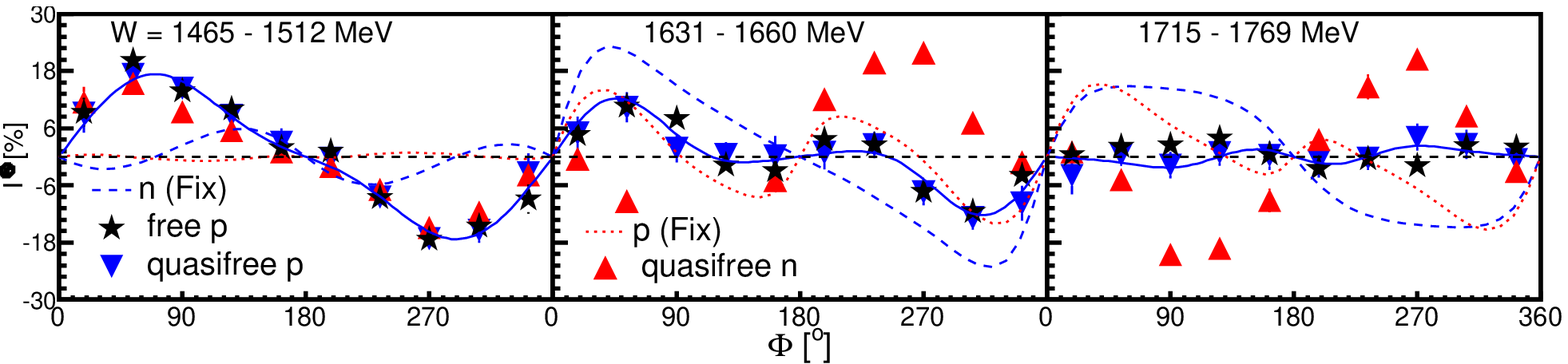}
      \put(40,3){{\color{green} \large{\rotatebox{45}{Preliminary}}}}
    \end{overpic}
  }}\newline
\vspace{-3mm}
\caption{
Top and middle row: $m(\gamma\gamma)$ and $\Delta M$ for the $\pi^{-} \pi^0 p$ final state, red 
triangles: data, vertical lines: applied cuts, dashed curve: background fit (only for 
$m(\gamma\gamma)$), black dashed line: MC for the $\pi^{-} \pi^0 p$ final state, cyan short 
dashed line: MC of $\pi^+ \pi^- \pi^0$ final states, blue full line sum of MC (only for $\Delta M$).
bottom row: Beam helicity asymmetry $I^{\odot}(\Phi)$ for the free (black stars) and quasi-free 
proton data (blue triangles) and for the quasi-free neutron data (red triangles) for the mixed-charge 
final states compared to model calculations from \protect\cite{Fix_05} (all results preliminary).
\label{fig:ana}
}
\end{figure}
The charged pions and the photons from the $\pi^0$ decay were detected in coincidence with the recoil 
nucleons.
Standard analysis steps, like cuts on the invariant mass m($\gamma\gamma$), 
missing mass $\Delta M =\ \left | P_{\gamma}\ +\ P_N\ -\ P_{\pi^0}\ -\ P_{\pi^{\pm}}\right | - m_N$ 
and coplanarity $\Delta \phi$ (azimuthal angular difference between the three-momenta of the two-pion 
system and the recoil nucleon), led to a very clean signal extraction, see Fig. \ref{fig:ana}. 

The beam-helicity asymmetry is defined as in Refs.~\cite{Roca_05,Oberle_13,Krambrich_09}: 
$I^{\odot}(\Phi)=\frac{d\sigma^{+}-d\sigma^{-}}{d\sigma^{+}+d\sigma^{-}}$
%=\frac{1}{P_{\gamma}}\frac{N^{+}-N^{-}}{N^{+}+N^{-}}$ 
where $d\sigma^{\pm}$ are the 
differential cross sections for each of the two photon helicity states.

A comparison of our free proton data
to previously measured beam-helicity asymmetries \cite{Krambrich_09} showed a good agreement.
In order to eliminate effects from the nuclear Fermi motion in the quasi-free data an analysis was performed
that used the invariant mass $W$ of the final state extracted from the four-momenta of the mesons and the recoil 
nucleons (rather than being approximated from the incident photon energy neglecting the Fermi momenta). 
The quasi-free proton results obtained this way are in good agreement with the free proton results.

\section{Results and conclusions}

The agreement between the free and quasi-free proton data demonstrates that the kinematic reconstruction 
of the final state eliminates effects from nuclear Fermi motion and that no other nuclear effects like
final state interactions are important.
%\vspace{-7mm}
\begin{figure}[!ht]
\centering{  \subfloat{%
    \begin{overpic}[width=0.388\textwidth]{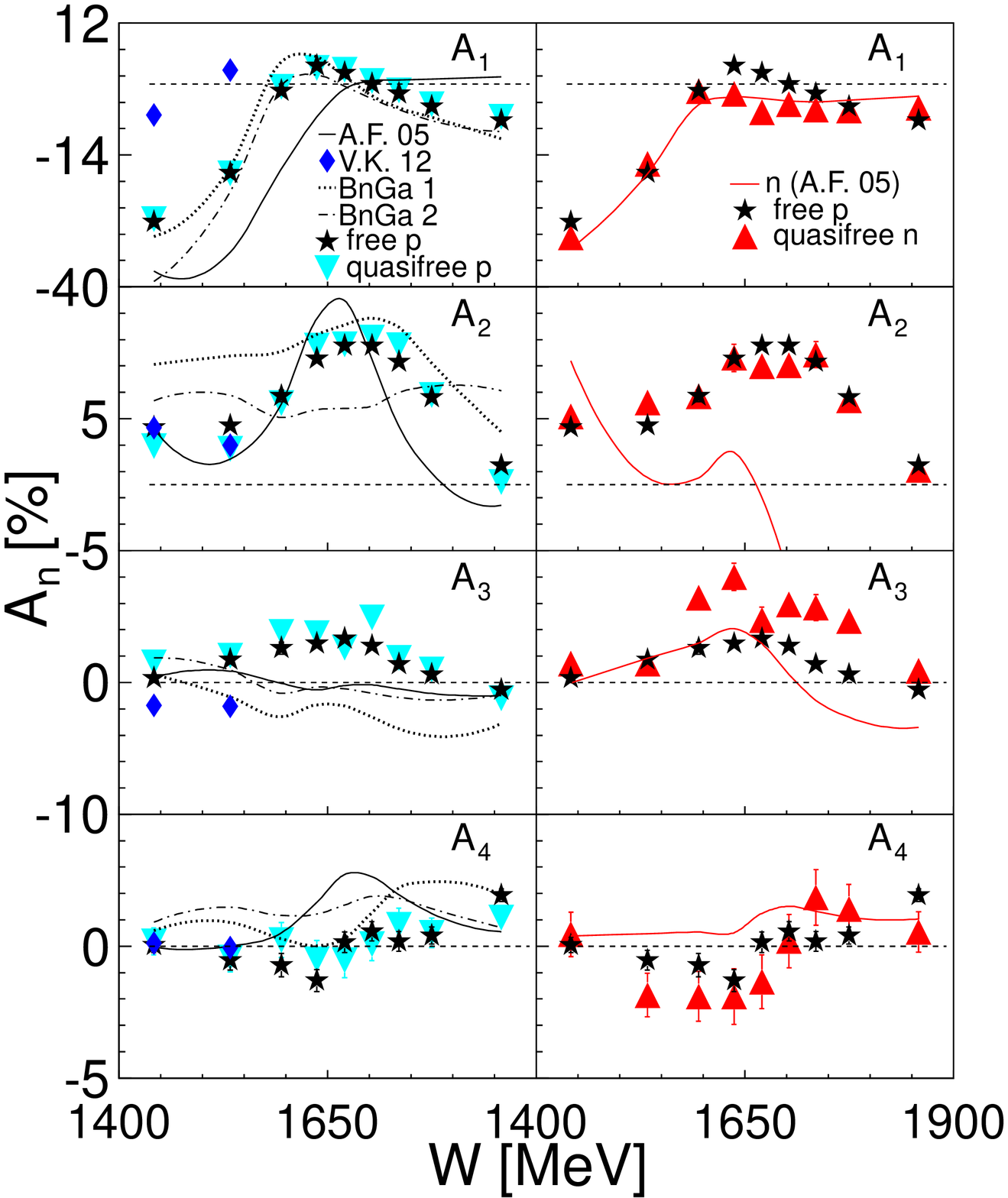}
      \put(30,30){}
    \end{overpic}
  }
  \subfloat{%
    \begin{overpic}[width=0.388\textwidth]{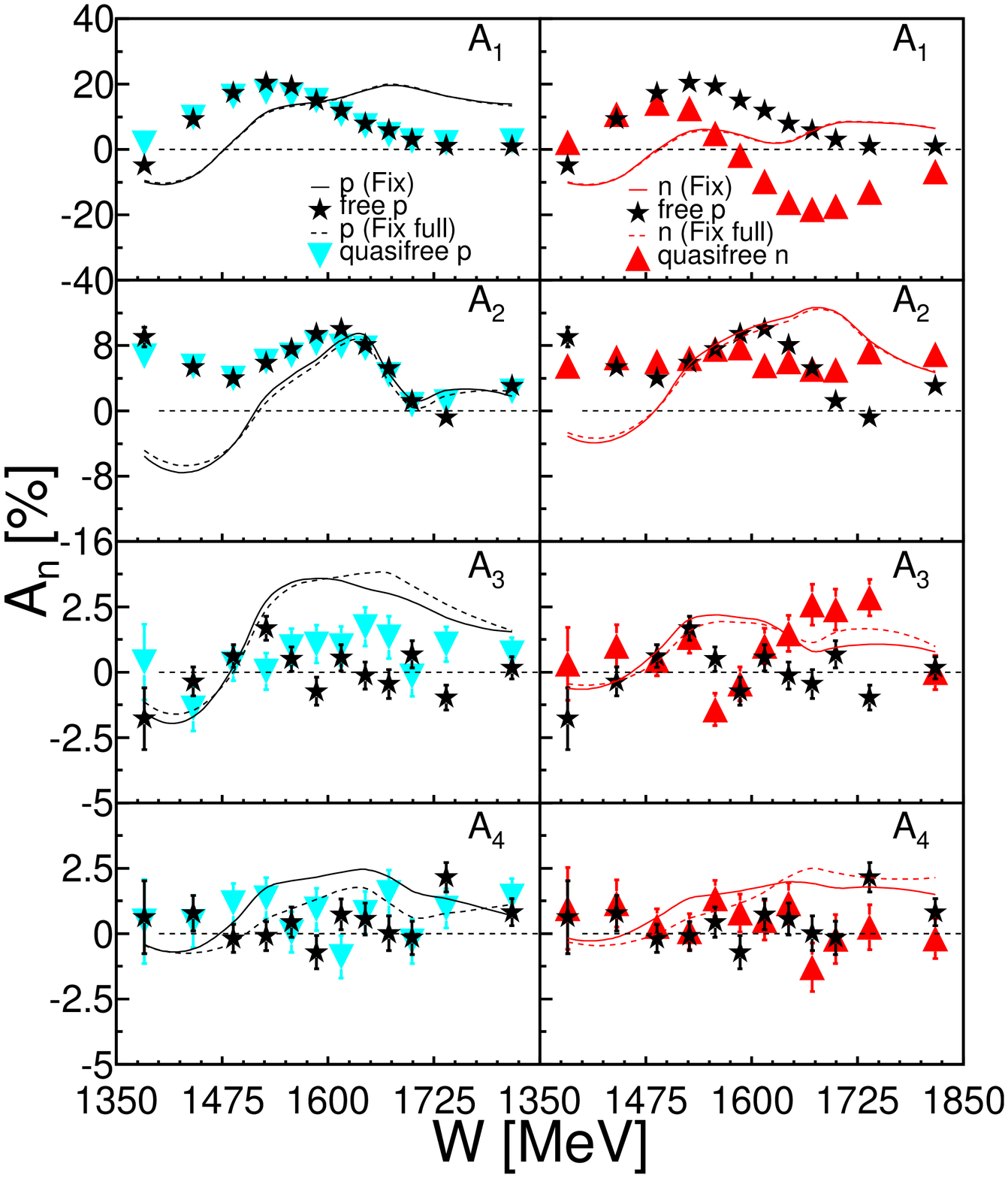}
      \put(23,33){{\color{green} \huge{\rotatebox{45}{Preliminary}}}}
    \end{overpic}
  }
\vspace{-1mm}
 \caption{left picture: coefficients of the fits to $I^{\odot}(\Phi)$ as function of the cm-energy $W$ for 
 the neutral channel; left-hand side: free and quasi-free proton data, right-hand side: comparison of proton 
 and neutron data. Solid curves: model results from \protect\cite{Fix_05}. Dotted curves: solution BnGa2011-1
 from \protect\cite{Anisovich_12}, dash-dotted curves: solution BnGa2011-2 from \protect\cite{Anisovich_12}.
 (Blue) diamonds: model results from \protect\cite{Kashevarov_12}.\newline right picture: coefficients for the 
 mixed-charge channel; same notation as above, model calculations from \protect\cite{Fix_05}
 \label{fig:I}}
}
\end{figure}
This suggests also that the quasi-free neutron data are a good approximation of a free neutron target.

Due to its symmetry properties, $I^{\odot}$ can be expanded in a sine series ($I^{\odot}(\Phi)=
\sum_{n=1}^{\infty}A_{n}\mbox{sin}(n\Phi)$) which can be fitted to the data. This allows for a more 
compact presentation of the results. In Fig. \ref{fig:I} the energy dependence of the parameters 
$A_n$ for $n\leq$4 (higher orders were not significant) are shown for free and quasi-free data and all four channels. 
The model calculations show a good agreement for $A_1$ of the neutral channel, whereas the agreement is less
good for $A_2$. For the mixed-charge asymmetries the agreement is rather poor and it seems that the
reaction mechanisms are still not understood very well. The present data will certainly constrain future 
model analyses of this reactions.

\footnotesize{

\vspace{2mm}
{\bf Acknowledgments}

We wish to acknowledge the outstanding support of the accelerator group 
and operators of MAMI. 
This work was supported by Schweizerischer Nationalfonds, Deutsche
Forschungsgemeinschaft (SFB 443, SFB/TR 16), DFG-RFBR (Grant No. 05-02-04014),
UK Science and Technology Facilities Council, STFC, European Community-Research 
Infrastructure Activity (FP6), the US DOE, US NSF and NSERC (Canada).

}


\begin{thebibliography}{0}    %for 1 digit

%%journal paper
% \bibitem{jpap} R. Loren and D. B. Benson, {\it J. Comput. 
% System Sci.} {\bf 27}, 400 (1983).

\bibitem{Krusche_03}            B. Krusche and S. Schadmand,                      Prog. Part. Nucl. Phys.           {\bf  51},      399    (2003). 
\bibitem{Sarantsev_08}          A.V. Sarantsev et al.,                            Phys. Lett.                     B {\bf 659},       94    (2008).
\bibitem{Thoma_08}              U. Thoma et al.,                                  Phys. Lett.                     B {\bf 659},       87    (2008).
\bibitem{Zehr_12}               F. Zehr et al.,                                   Eur. Phys. J.                   A {\bf  48},       98    (2012).
\bibitem{Kashevarov_12}         V. Kashevarov et al.,                             Phys. Rev.                      C {\bf  85},   064610    (2012).
\bibitem{Ajaka_08}              J. Ajaka et al.,                                  Phys. Rev. Lett.                  {\bf 100},   052003    (2008).
\bibitem{Horn_08a}              I. Horn et al.,                                   Phys. Rev. Lett.                  {\bf 101},   202002    (2008).
\bibitem{Horn_08b}              I. Horn et al.,                                   Eur. Phys. J.                   A {\bf  38},      173    (2008).
\bibitem{Kashevarov_09}         V. Kashevarov et al.,                             Eur. Phys. J.                   A {\bf  42},      141    (2009).
\bibitem{Kashevarov_10}         V. Kashevarov et al.,                             Phys. Lett.                     B {\bf 693},      551    (2010).  
\bibitem{Roberts_05}            W. Roberts and T. Oed,                            Phys. Rev.                      C {\bf  71},   055201    (2005).
\bibitem{Herminghaus_83}        H. Herminghaus et al.,                            IEEE Trans. on Nucl. Science.     {\bf  30},     3274    (1983).
\bibitem{Kaiser_08}             K.-H. Kaiser et al.,                              Nucl. Instr. Meth.              A {\bf 593},      159    (2008).
\bibitem{Anthony_91}            I. Anthony et al.,                                Nucl. Instr. Meth.              A {\bf 301},      230    (1991).
\bibitem{Hall_96}               S.J. Hall, G.J. Miller, R. Beck, P.Jennewein,     Nucl. Inst.Meth.                A {\bf 368},      698    (1996).
\bibitem{McGeorge_08}           J.C. McGeorge et al.,                             Eur. Phys. J.                   A {\bf  37},      129    (2008).. 
\bibitem{Starostin_01}          A. Starostin et al.,                              Phys. Rev.                      C {\bf  64},   055205    (2001).
\bibitem{Novotny_91}            R. Novotny,                                       IEEE Trans. on Nucl. Science      {\bf  38},      379    (1991). 
\bibitem{Gabler_94}             A.R. Gabler et al.,                               Nucl. Instr. Meth.              A {\bf 346},      168    (1994).
\bibitem{Schumann_10}           S. Schumann et al.,                               Eur. Phys. J.                   A {\bf  43}       269    (2010). 
\bibitem{Watts_04}              D. Watts, in {\em Calorimetry in Particle Physics, Proceedings of the 11th International Conference, Perugia, Italy 2004}, edited 
by C. Cecchi, P. Cenci, P. Lubrano, and M. Pepe (World Scientific, Singapore, 2005), p. 560
\bibitem{Fix_05}                A. Fix and H. Ahrenh\"ovel,                       Eur. Phys. J.                   A  {\bf 25}       115    (2005).
\bibitem{Roca_05}               L. Roca,                                          Nucl. Phys.                     A {\bf 748},      192    (2005). 
\bibitem{Oberle_13}             M. Oberle et al.,                                 Phys. Lett.                     B {\bf 721},      237    (2012).
\bibitem{Krambrich_09}          D. Krambrich et al.,                              Phys. Rev. Lett.                  {\bf 103},   052002    (2009).
\bibitem{Anisovich_12}          A.V. Anisovich et al.,                            Eur. Phys. J.                   A  {\bf 48}        15    (2012).
\end{thebibliography}
\end{document}